\documentclass[lettersize,journal]{IEEEtran}
\usepackage{amsmath,amsfonts}
\usepackage{algorithmic}
\usepackage{algorithm}
\usepackage{array}
\usepackage{textcomp}
\usepackage{stfloats}
\usepackage{url}
\usepackage{verbatim}
\usepackage{cite}

\usepackage{pifont}
\usepackage{subcaption}
\usepackage{hyperref}
\usepackage{mathtools}
\usepackage[skip=5pt]{caption}
\usepackage{fancyhdr}
\usepackage{multirow}
\usepackage{ragged2e} 
\usepackage{booktabs,makecell, multirow, tabularx}
\begin{document}

\title{LTRDetector: Exploring Long-Term Relationship \\ for Advanced Persistent Threats Detection}


\author{Xiaoxiao Liu$^{\dagger}$, Fan Xu$^{\dagger}$, Nan Wang$^{*}$, Qinxin Zhao, Dalin Zhang, Xibin Zhao$^{*}$, Jiqiang Liu
\thanks{$^{\dagger}$ Equal contributions.}
\thanks{$*$ Nan Wang and Xibin Zhao are the corresponding authors.}
}




\maketitle

\begin{abstract}
    Advanced Persistent Threat (APT) is challenging to detect due to prolonged duration, infrequent occurrence, and adept concealment techniques. Existing approaches primarily concentrate on the observable traits of attack behaviors, neglecting the intricate relationships formed throughout the persistent attack lifecycle. Thus, we present an innovative APT detection framework named LTRDetector, implementing an end-to-end holistic operation. LTRDetector employs an innovative graph embedding technique to retain comprehensive contextual information, then derives long-term features from these embedded provenance graphs. During the process, we compress the data of the system provenance graph for effective feature learning. Furthermore, in order to detect attacks conducted by using zero-day exploits, we captured the system's regular behavior and detects abnormal activities without relying on predefined attack signatures. We also conducted extensive evaluations using five prominent datasets, the efficacy evaluation of which underscores the superiority of LTRDetector compared to existing state-of-the-art techniques.
\end{abstract}


\section{Introduction}
\IEEEPARstart{A}{dvanced} Persistent Threat (APT) has become one of the most critical cyberspace threats to enterprises and institutions\cite{alshamrani2019survey}, which results in significant financial losses. In one of the first detailed reports on Advanced and Persistent Threats (entitled APT1\cite{Mandiant}), the security firm Mandiant disclosed the goals and activities of a global APT actor. The activities contained stealing hundreds of terabytes of sensitive data (including business plans, technology blueprints, and test results) from at least 141 organizations across a diverse set of industries. They estimated that malware would remain in target organizations for an average of 365 days before being discovered, which suggested that APT attacks are frequently covert and long-lasting. In addition, the adversary of an APT often leverages zero-day exploits\cite{2011APT, virvilis2014big, sun2018using} to take over a system and continuously listen to it for an extended period\cite{2020Trustwave}. Due to these characteristics of APT attacks, traditional security tools struggle to effectively detect and defend against them.

Nowadays, more and more academics have focused on provenance graph-based attack detection\cite{milajerdi2019holmes, han2020unicorn}. However, traditional detection systems are not suitable for APT. Some of these approaches\cite{koren2009matrix, shervashidze2011weisfeiler} are incapable of modeling runtime APT attacks, and the loss of information in the feature representation rises as the data grows. Meanwhile, APT attack has the characteristics of long duration, high stealth, and low frequency. It generally takes complex long-period attacks compared with traditional attacks, and if individual attack steps are buried in the background "noise" of normal behavior, it is not possible to effectively identify it. As a result, it is crucial for APT detection methods to accumulate low-frequency anomalous behavior to generate long-term features. Furthermore, most current approaches\cite{hassan2019nodoze, xie2018pagoda} rely on rule design and expert experience, and cannot automatically extract normal or anomalous features from the data, making them incapable against APT attacks such as zero-day exploits.

\begin{figure*}[t]  
    \centering
    \includegraphics[width=0.9\linewidth]{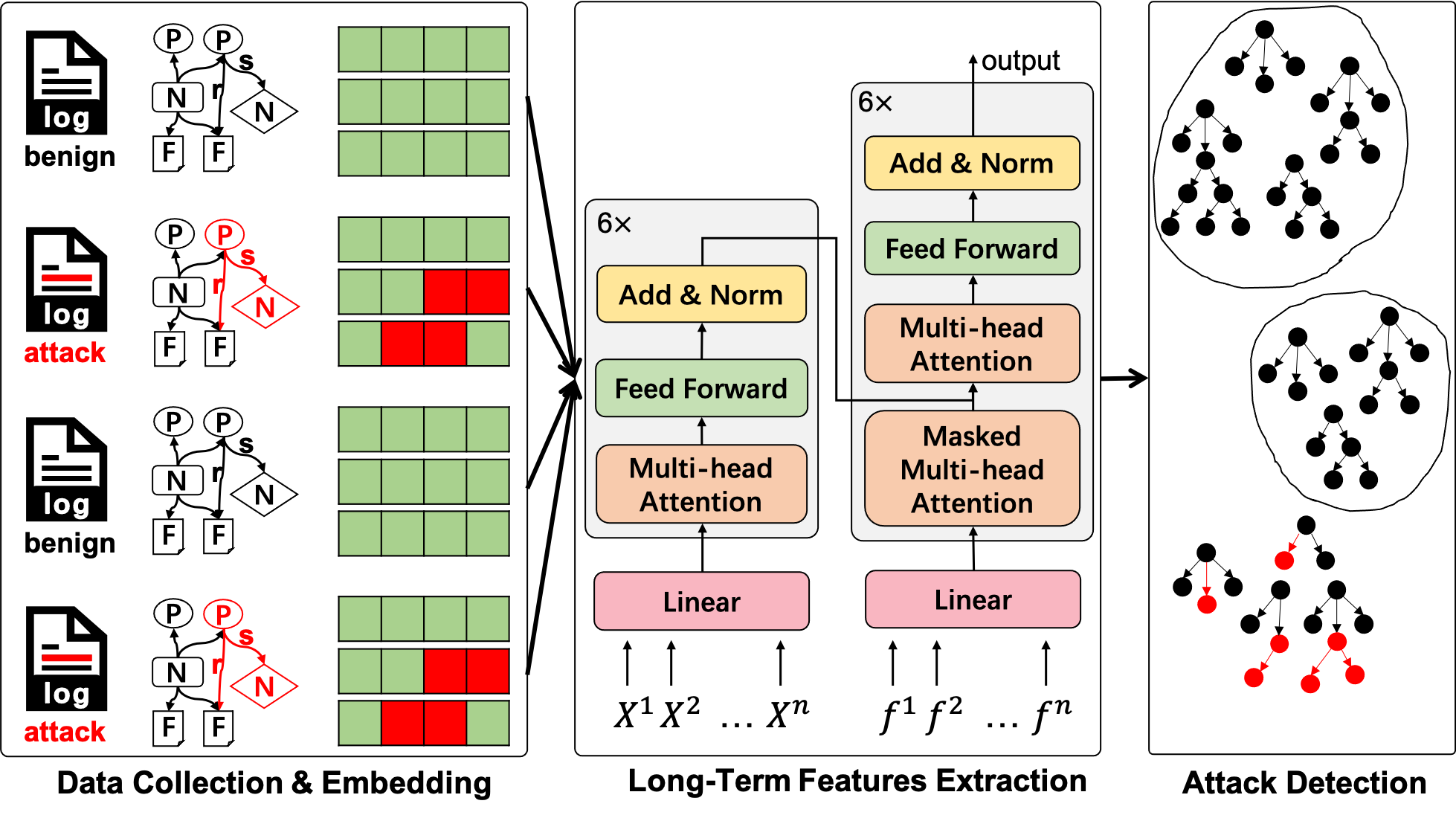}
    \caption{The framework of {\it LTRDetector}.}\label{framework}
\end{figure*}

To tackle these issues, we propose {\it LTRDetector}, a provenance graph-based APT attack detector. It analyzes provenance graph data using a novel deep learning approach that does not require prior knowledge of attack behavior, eliminating the need for manual analysis to label training sets. As shown in Figure\ref{framework}, the framework of our method consists of three stages. More specifically, in the first phase, we present a graph-based real-time APT attack activity embedding approach that effectively reduces redundant data while generating a set of feature sequences with rich context information. In the second stage, {\it LTRDetector} developed an Autoencoder structure model based on the multi-head attention algorithm to extract the long-term features of the provenance graph and obtain the long-term correlation of system behavior, aiming at the characteristics of long latency and strong concealment of APT attack. In the last phase, we employ a clustering analysis algorithm to model the behavior of the system during training, and identify any behavior that exceeds a predefined threshold as an attack activity, achieving unsupervised learning. \\The main contributions of our paper are as follows: 
\begin{itemize}
    \item We present a novel approach for provenance graph representation that can effectively handle the embedding of low-frequency information in a large amount of data and it also can adapt to changing network structures without the need for a full recalculation because it produces a good intermediate representation.
    \item To efficiently mine the potential association relationships generated by persistent APT attacks, we employ an efficient long-term feature extraction method that can extract the full association relationships in the sequence without a long-range dependence problem.
    \item Considering the immediacy of APT attacks and the need for extensive manual analysis to label attack data. {\it LTRDetector} proposes a method that can identify anomalous activities without predefined attack features, eliminating the need for human analysts. 
    \item We compare our method with state-of-the-art methods on five widely used datasets and the experimental findings show that {\it LTRDetector} can detect real-life APT scenarios with high accuracy.
\end{itemize}

The remainder of this article is organized as follows. 
In Section\ref{related} we introduce the recent research status of APT attack detection based on the provenance graph. Section\ref{LTRDetector}  shows the overall structure of our method and introduces each step in detail. In Section\ref{training}, we introduce the model training process. In Section\ref{experiment}, we give the experimental results of our detection method compared with the current three state-of-the-art methods. Finally, we conclude our paper in Section\ref{conclusion}.

\section{Related Work}\label{related}
In recent years, provenance graph has been widely used in attack detection and restoration\cite{milajerdi2019holmes,han2020unicorn}, since they store a wealth of information about the system’s activity. APT attack detection based on provenance graphs can be divided into three categories: anomaly-score-based attack detection\cite{hassan2019nodoze,xie2018pagoda}, tag-propagation-based attack detection\cite{milajerdi2019holmes,hossain2017sleuth}, and graph-matching-based attack detection\cite{han2020unicorn}. 

For the first type, anomaly-score-based attack detection methods consider the anomalous score of a single provenance path or the entire provenance graph. For example, Xie \emph{et al.} established the normal rule base in the paper\cite{2016Unifying,xie2018pagoda} and the abnormal rule base in the paper\cite{Gaussian}. They proposed a method named Pagoda\cite{xie2018pagoda}. This approach computes the anomaly score of each path, then multiplies the length of each path by its anomaly score to generate a weight value, and finally computes the overall provenance graph's anomaly score. When the score of anomaly exceeds a predefined threshold, the system is judged to have been attacked. NoDoze\cite{hassan2019nodoze} presented that threat detection softwares generate more alerts than cyber analysts can properly investigate. This leads to a "threat alert fatigue" or information overload problem where cyber analysts miss true attack alerts in the noise of false alarms. They present NoDoze to eliminate alarm fatigue by calculating the anomaly score of the path on the provenance graph. Sun \emph{et al.}\cite{2016Towards, 2018Using} suggested modeling the provenance graph using a Bayesian network. To determine the zero-day attack, the suspicious information is first mapped to the nodes in the Bayesian network, and then the conditional probability table is used to assess the likelihood that other nodes will become infected. Nevertheless, the methods described above are very time-consuming in calculating the anomaly score of the entire provenance graph and they cannot automatically extract normal or anomalous features from the data since they rely on rule design and expert knowledge.

Tag-propagation-based attack detection is another popular strategy. As the name implies, this strategy presupposes that labeled data are sufficient to extract valuable information. Hossain \emph{et al.} proposed SLEUTH\cite{hossain2017sleuth}, which uses provenance graphs to restore real-time attack scenarios. SLEUTH starts from anomalous nodes, performs backward analysis, finds the entry node with entry degree 0 using the single source shortest path algorithm, starts from the entry point, and performs forward search and pruning based on the path weights, and the search results. In order to address the issue that the graph produced by SLEUTH contains a lot of benign nodes when dealing with Long-Term attacks. Hossain \emph{et al.}\cite{hossain2020combating} introduced two attenuation strategies in the tag propagation process, lowering the number of benign nodes identified as suspicious nodes and resolving the dependency explosion issue. Even though the provenance graph has a wealth of system-level data, higher-level attack semantic data is missing. To address this issue, Milajerdi \emph{et al.}\cite{milajerdi2019holmes} developed a set of mapping rules to map a provenance graph with low-level information to a chain model with higher-level semantics, which created an advanced scenario graph that serves as the basis for detecting attacks. These techniques can detect APT attacks, but they necessitate domain expertise and expert experience, and their efficacy depends on the rule design.

The third form, graph-matching attack detection has grown in popularity during the last few years. Milajerdi \emph{et al.} proposed POIROT\cite{milajerdi2019poirot}, and attacks are discovered as subgraphs in the provenance graph that are similar to the attack graph. Although the graph matching-based approach detects attacks more precisely, it necessitates the creation of attack graphs based on prior knowledge and cannot detect unknown attacks. These systems focus on rule design and expert experience, rather than learning normal or attack patterns from data. The similarity of local subgraphs of a provenance graph\cite{manzoor2016fast} was used to define the similarity between provenance graphs, and similar provenance graphs were clustered together by a clustering algorithm. However, the similarity of the entire provenance graph could not reflect the attack temporality. UNICORN\cite{han2020unicorn} first transforms the provenance graphs into feature vectors and clusters the feature vectors to generate system states, then it uses an automaton to describe system state changes. But UNICORN still has several issues, for APT attacks are persistent and covert, and there is little distinction between a single feature vector and an anomalous behavior feature vector in long sequences. So it is unable to accumulate these differences and accurately identify real-time situations.  Liang \emph{et al.} proposed SeqNet (Sequence Networks\cite{2022SeqNet}), which is a deep learning-based attack detection technique on provenance graphs. SeqNet converts provenance graphs into feature vectors using the same technique as UNICORN, and then they use the GRU model to extract Long-Term features of the provenance graph sequence. Nevertheless, the GRU, as we know it as an RNN model with typical recurrent neural network and long short-term memory model features, cannot solve the Long-Term dependency problem\cite{bengio1994learning,wu2024spatio}, in which past information is retained through hidden states, but the earlier information recorded in the memory unit is washed out over time steps, making it impossible to establish a relationship with the earlier time information dependencies. 

Consequently, prior research explored the use of data provenance to detect APT attacks but failed to do so effectively and accurately for ignoring the persistence and low frequency of APT attacks. Furthermore, the generation of node tags and the statistical data in the form of a histogram increases linearly with the amount of training data collected. As a result, information loss increases, as does detection time.

\section{LTRDetector}\label{LTRDetector}
{\it LTRDetector} detects APT attacks by searching anomalous system behaviors. Its operation consists of three stages: \ding{172} data embedding, \ding{173} Long-Term features extraction, and \ding{174} attack detection. The architecture and workflow of {\it LTRDetector} are depicted in Figure\ref{framework}.

More specifically, the first step is data embedding. To start, programs like CamFLow\cite{pasquier2017practical} will gather the system log and create a trace graph that includes every system call step. The provenance graph is then compressed using the CPR\cite{2016High} method. We next apply the BFS random walk to capture the topology information of the provenance graph in parallel, and employ a graph embedding algorithm to characterize provenance graph nodes. The second step is long-term features extraction. After generating a feature sequence describing system changes, we introduce a Transformer-based Multi-head attention\cite{vaswani2017attention} network to extract the long-term features of the provenance graphs. The last step is attack detection. We use the clustering algorithm to divide the training set features into K classes as K clustering centers. In the attack detection phase, we extract the long-term features of each provenance graph using the model's Encoder and calculate the distance between the extracted feature vector and the cluster center, and we classify any behavior that exceeds the predetermined threshold as an attack activity.

\begin{figure}[t]  
    \centering
    \includegraphics[width=3in,height=1.4in]{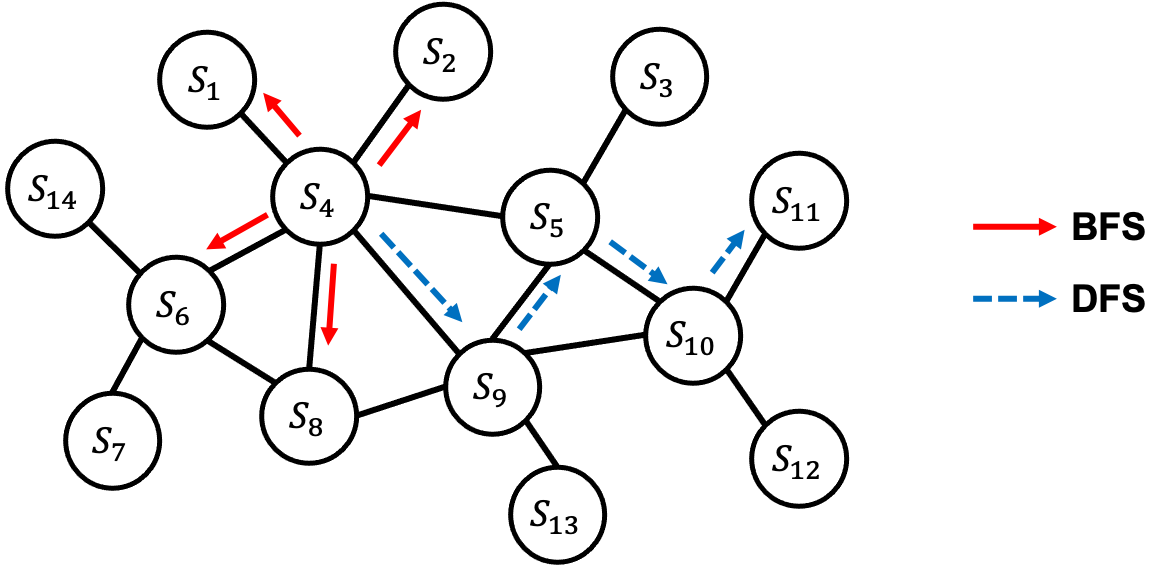}
    \caption{An example of the BFS and DFS random walk.}\label{BFS}
\end{figure}

\subsection{Data embedding}
{\it LTRDetector} accepts a stream of attributed edges generated by a provenance graph capture system that is running on one or more networked hosts. A single, whole-system provenance Directed Acyclic Graph (DAG) is created by provenance systems, and a partial-order guarantee makes it possible to do fully contextualized analysis and streaming computation effectively. {\it LTRDetector} selects CamFlow\cite{pasquier2017practical} as its provenance construction system, and alternative systems such as LPM and Spade\cite{gehani2012spade} were also viable alternatives.

A set of informative, discriminative, and independent characteristics are necessary for feature learning\cite{grover2016node2vec,wu2023pastnet}. Node featurization is a crucial component of graph embedding. There are two key requirements that an effective embedding method for system-level provenance graph nodes must meet. First, system entities play a specific role in the system, so their embeddings must be close if their roles are similar. Second, the embedding strategy needs to retain as much semantic information as possible. We successfully meet both of these requirements by effectively employing the breadth-first random walk and  Skip-gram\cite{mikolov2013distributed} model to characterize the contextual information of each node in an embedding space. Due to the characteristics of the provenance graph collection tool, the subject and object are versioned. This will significantly increase the quantity of the provenance graph node, so version control can be optimized by deleting the majority of the redundant events in the log without modifying the provenance graph causality. We compress the data of the system provenance graph with CPR (Causality-Preserving Reduction) and FDR (Full Dependence-Preserving Reduction)\cite{li2021threat}. These two compression algorithms can eliminate repetitive actions without modifying the causal relationship between objects. In addition, removing duplicate edges makes it easier to extract the nodes' structural information for the ensuing node embedding.

Figure\ref{BFS} shows that a BFS random walk is more likely to learn structural equivalence characteristics compared to a DFS random walk. Therefore, this paper uses BFS random walk to extract the local topology information of the network. In a provenance graph, each node, whether file or process, corresponds to a series of walk paths that encode valuable contextual information. To build the causal context for each node, {\it LTRDetector} conducts directed BFS random walks of a fixed length ${\boldsymbol{l}}$. The {\it LTRDetector} traverses the graph by following edges when given a source node, such as $c_0$. If a node has many outgoing edges, {\it LTRDetector} chooses one at random to proceed with the walk. The causal context C for ${c_0}$ is \{ ${c_i|}$i = 1,\ldots,  ${\boldsymbol{l}}$\}, where ${c_i}$ is generated by the distribution:
\begin{equation}
P(c_i=v|c_{i-1}=u)= 
\begin{cases}
    \frac{1}{N} & \mbox{if }{(u,v) \in E} \\
    0 & \mbox{otherwise }
\end{cases}, 
\end{equation} 
where $N$ is the number of outgoing edges from $c_{i-1}$, and (u,v) is one of the directed edges.

\begin{figure}[t]  
    \centering
    \includegraphics[width=3.2in,height=2in]{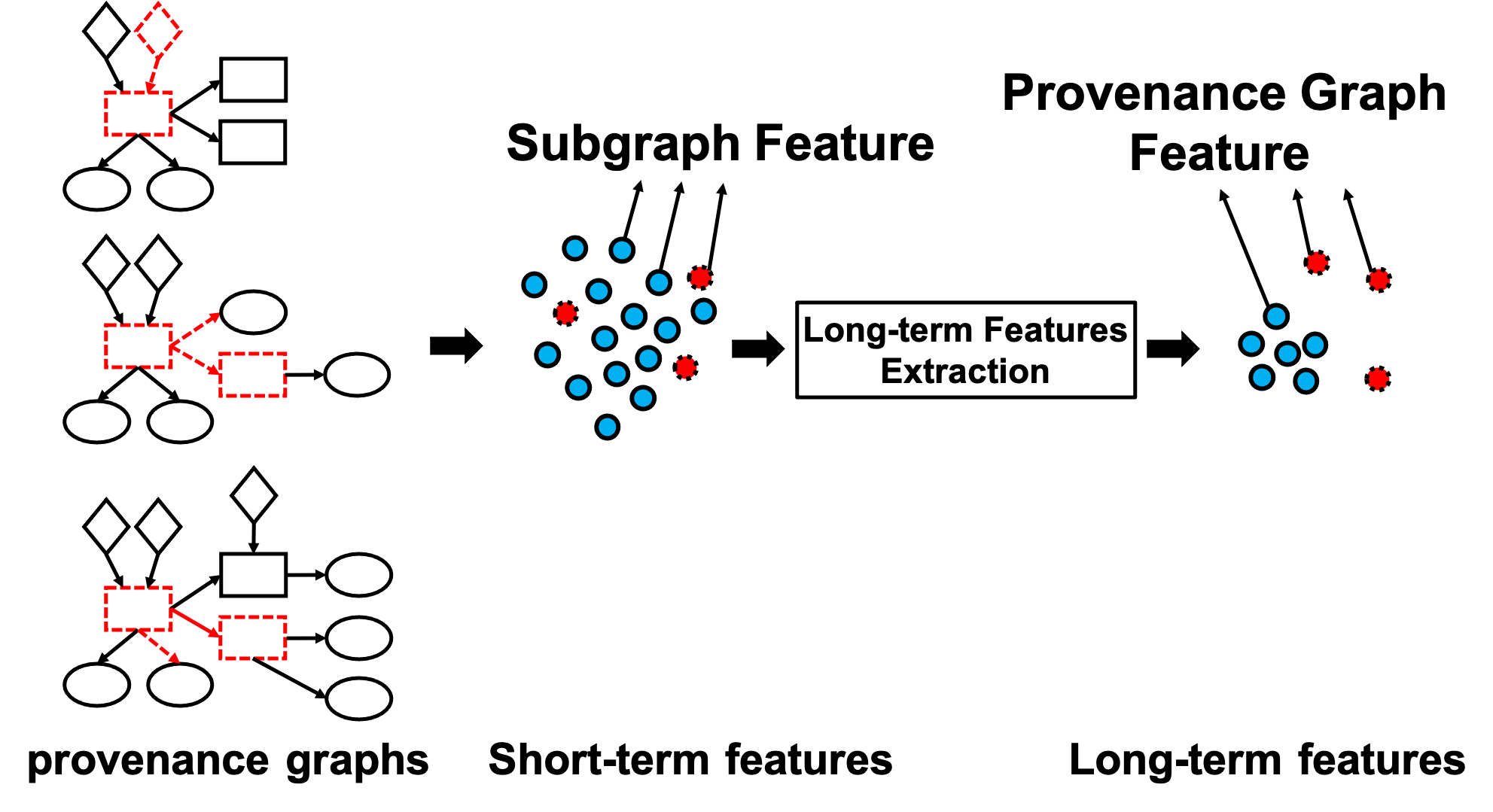}
    \caption{An example of low-frequency anomalous behaviors can accumulate continuously and reflect significant differences from normal behaviors.}\label{long-term}
\end{figure}

Then we employ the Word2Vec\cite{mikolov2013distributed,wu2024slfnet} algorithm as a foundation for our embedding approach to characterize provenance graph nodes. Compared to CBOW (Continuous Bag of Words), the objective of the skip-gram\cite{mikolov2013efficient} model is based on the distributional hypothesis which states that words in similar contexts tend to have similar meanings. That is, similar words tend to appear in similar word neighborhoods. In addition, skip-gram is better at capturing the meaning of rare words and phrases in larger vocabularies and is also more robust to noisy data whose characteristics are more consistent with long-duration and low-frequency APT attacks. Skip-gram embeds words into a low-dimensional continuous vector space, where words with similar context map closely together. Given a sequence of words, the Word2Vec utilizes the skip-gram model to optimize the log probability of predicting the context around a specific target word. The context is determined by a fixed-size sliding window over the text sequence. Assuming that given the target word, the likelihood of observing each context word is independent, the Word2Vec maximizes:
\begin{equation}
    \max \frac{1}{\mathrm{T}}\sum_{t=1}^T\sum_{-c \leq j \leq c,j \ne 0} \log \frac{\exp({v_{w_{t+j}}^{'}}^\mathsf{T}v_{w_t})}{\begin{matrix} \sum_{w=1}^W \exp({v_w^{'}}^\mathsf{T}v_{w_t})\end{matrix}}, 
\end{equation}
where c is the window size. $v_{w_{t+j}}$ and $v_{w_t}$ are the embeddings of the context word and the target word. $W$ is the vocabulary size.

\subsection{Long-Term Features Extraction}

\begin{figure}[t]  
    \centering
    \includegraphics[width=2.8in,height=2.2in]{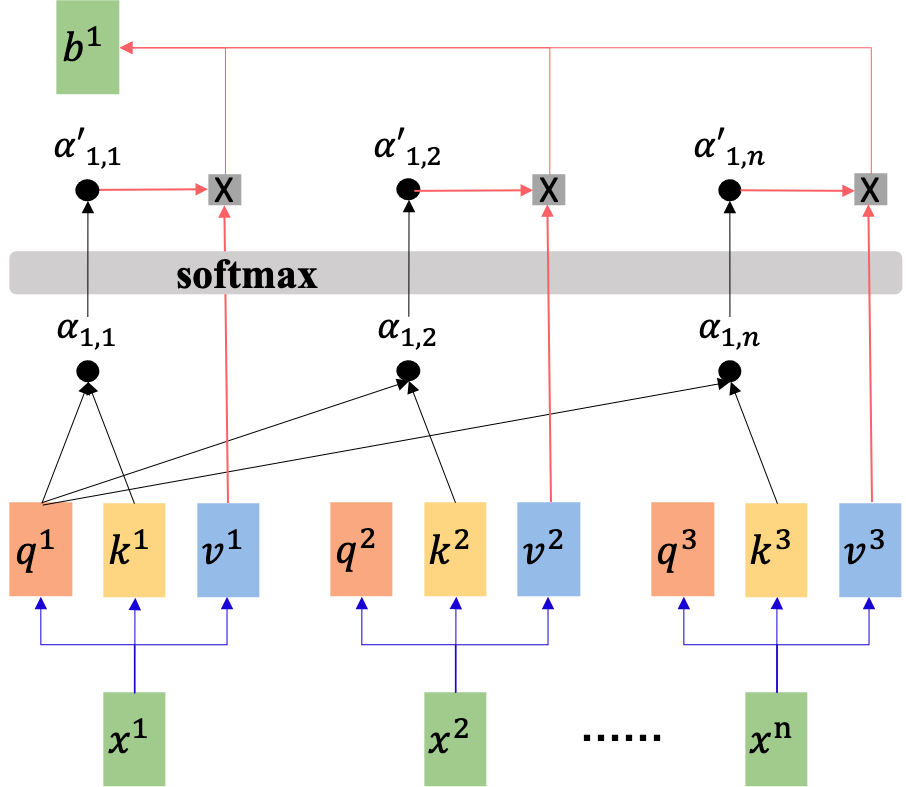}
    \caption{Long-Term features extraction.}\label{attention}
\end{figure}

APT attack has the characteristics of long duration, high stealth, and low frequency. As shown in Figure\ref{long-term}, each step of APT attack activity represents a small portion of the overall system activity, which makes it challenging to distinguish between normal behavior and abnormal behavior in the short term. But, anomalous behavior accumulates in the long term, resulting in significant differences from normal behavior. 

In response to this feature, detection approaches are necessary to capture long-term dependencies in system behavior that include long-term interactions between nodes. We investigated that attention mechanisms can effectively extract Long-Term features of sequences, which substitutes the cycling structure in the original Seq2Seq model by weighting each position in the input sequence. Since the self-attention module can establish a direct relationship between any two points in a sequence, it can well capture the correlation between long-period APT attack behaviors. In this paper, we apply this principle to model the system's state changes and use Multi-Head attention\cite{vaswani2017attention,yan2024dual} to extract Long-Term features of the provenance graph.

\subsubsection{Scaled Dot-Product Attention}

As shown in Figure\ref{attention}, it introduces three matrices Q, K, and V. Letters Q, K, and V stand for the query vector, keyword vector, and content vector, respectively. The dot product of Q and K yields the correlation matrix between vectors, then through the softmax gets the weight of attention. Finally, the weights are applied to V to get the final attention value, achieving the purpose of weighting different data. The following is the attention formula:

\begin{equation}
Attention(Q,K,V) = softmax(\frac{QK^\mathsf{T}}{\sqrt{d_k}})V, 
\end{equation}
where Q, K, and V stand for the query vector, keyword vector, and content vector; $d_k$ is the dimension of queries and keys.

There are two most commonly used attention functions: additive attention\cite{2014Neural}, and dot-product (multiplicative) attention. Additive attention computes the compatibility function using a feed-forward network with a single hidden layer, while dot-product attention can be implemented using highly optimized matrix multiplication code. We choose dot-product attention because it is much faster and more space-efficient in practice.

\subsubsection{Multi-Head Attention} 

This module extracts various features with different attention. The basic goal is to learn multiple representations by using different attention to extract different information, similar to how CNN uses a multi-channel convolution kernel to extract different features. This improves the model's capacity to generalize.

In order to better capture long-range dependencies in the sequence, Multi-Head attention\cite{vaswani2017attention} linearly projects the queries, keys, and values to $h$ times with different, learned linear projections to $d_k$, $d_k$, and $d_v$ dimension, where each dimension stands for a distinct representation subspace. On each of these projected versions of queries, keys, and values we then perform the attention function in parallel, yielding $d_v$-dimensional output values. These are concatenated and once again projected, resulting in the final values, as depicted in the following formula:
\begin{align}
& MultiHead(Q,K, V) = Concat(head_1,\ldots, head_h)W^O, \\
& where \ head_i = Attention(Q{W_i}^Q ,K{W_i}^K , V{W_i}^V), \nonumber 
\end{align}
Where the projections are parameter matrices ${W_i}^Q {\in} {{\mathbb R}^{d_{model} \times d_k}}$, ${W_i}^K {\in} {{\mathbb R}^{d_{model} \times d_k}}$, ${W_i}^V {\in} {{\mathbb R}^{d_{model} \times d_v}}$ and $W^O {\in} {{\mathbb R}^{hd_v \times d_{model}}}$.  

\subsubsection{Masked Self-attention} 

Self-attention in the model's Decoder requires some restrictions to ensure that the Decoder does not see future information, hence it is required to limit the $Q$ of each position to focus only on the current position and the previous position. We introduce a mask matrix to add to the attention score, where the illegal position is occluded by -$\infty$.

\subsubsection{Position-wise Feed-Forward Networks} 

Position-wise Feed-Forward Networks act on each module to improve the model's fitting capabilities. It is a multi-layer perceptron (MLP) with two layers: a fully connected layer and a nonlinear activation function. The following is the precise formula:

\begin{equation}
FFN(H) = ReLU(HW_1+b_1)W_2+b_2, 
\end{equation}
where H is the preceding layer's output. $W_1$, $W_2$, $b_1$, $b_2$ are trainable parameters.

\subsubsection{Residual and Normalization}

Residual\cite{2016Deep,wu2023spatio} and normalization\cite{2016Layer} solve the problems of gradient disappearance, gradient explosion, and network degradation in deep network training. We apply dropout\cite{2014Dropout} to the output of each sub-layer, before it is added to the sub-layer input and normalized. In addition, we apply dropout to the sums of the embeddings and the positional encodings in both the encoder and decoder stacks. In each layer:

\begin{align}
& H \ = LayerNorm(SelfAttention(X)+X), \\
& H^{'} = LayerNorm(FFN(H)+H), 
\end{align}
where $X$ is the input vector, and $H$ is the preceding layer's output.

\subsection{Attack detection}

Models trained based on current attack behaviors cannot successfully detect novel attack techniques because of the diversity and unpredictability of APT attacks. As can be seen, if the existing attack data is used to train the model in the APT attack detection scenario, it will only be able to identify existing attack techniques and not be able to cover new attack techniques. Additionally, the model is easily overfitting, which reduces its ability to generalize to unknown attacks.

To address this situation, this paper uses the clustering algorithm K-means\cite{1979Algorithm} to group all the feature vectors collected from the training set into K classes. In the attack detection phase, we only use the Encoder of the model to extract the long-term feature of the input sequence. These feature vectors contain system behavior characteristics, so the behaviors with similar attributes will be adjacent to one another in the embedding space. According to this feature, we calculate the distance between the extracted feature vector and the cluster center in the attack detection stage, and any behavior that exceeds the threshold is considered an attack.

\begin{figure}[t]  
    \centering
    \includegraphics[width=2.6in,height=3.2in]{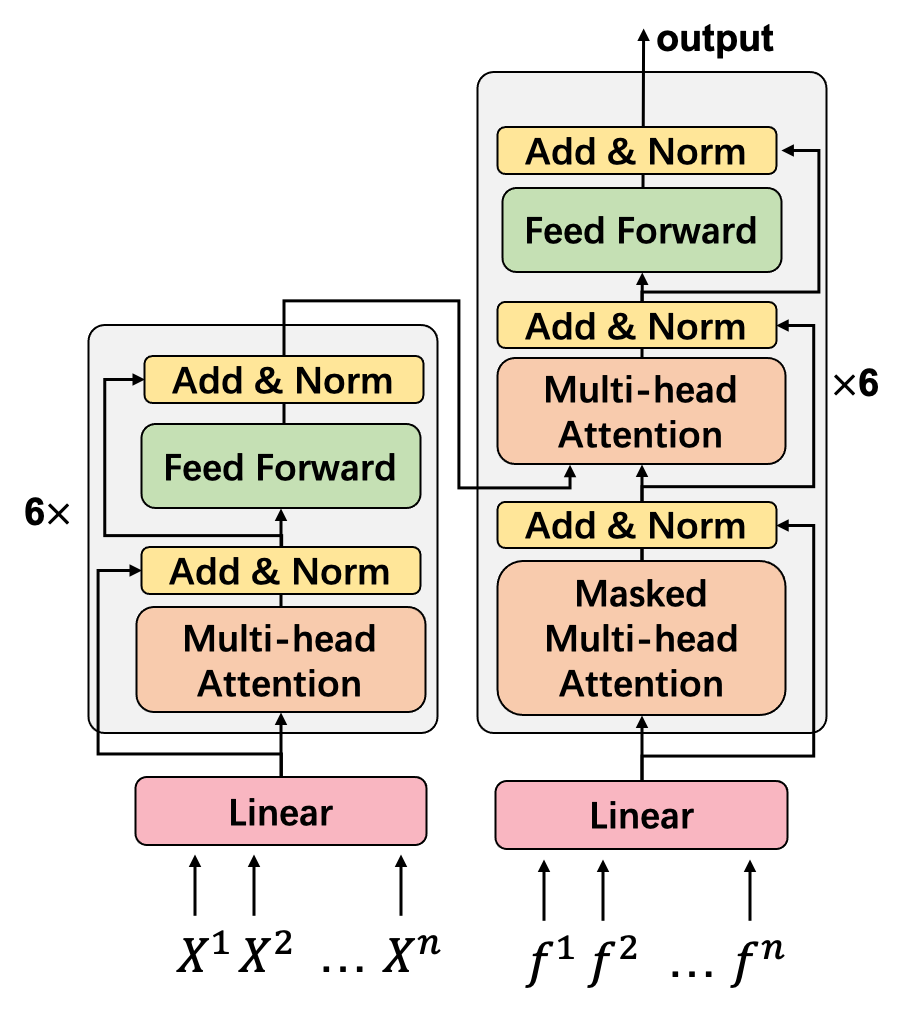}
    \caption{The Long-Term features extraction model. The input \{${X^1,X^2,\ldots,X^n}$\} is the feature sequence of a provenance graph generated by Word2Vec in the precivious step. The input \{${f_1,f_2,\ldots,f_n}$\} is the feature vector of a provenance graph generated by the Encoder.}\label{encoder_decoder}
\end{figure}

\section{Model training}\label{training}
Figure\ref{encoder_decoder} shows that our model is an encoder-decoder structure and it follows this overall architecture using stacked Multi-head attention\cite{vaswani2017attention} and Position-wise Feed-Forward for both the encoder and decoder.
\begin{itemize}
    \item Encoder: The encoder is composed of a stack of $N = 6$ identical layers. Each layer has two modules. The first is a Multi-head self-attention mechanism, whose head number is set $h=8$ in our model, and the second is a position-wise feed-forward network. We employ a residual connection\cite{2016Deep} around each of the two sub-layers, followed by LayerNorm\cite{2016Layer}. The output of each sub-layer is $LayerNorm(X + Sublayer(X))$.
    \item Decoder: The decoder is also composed of a stack of $N = 6$ identical layers. In addition to the two modules in each encoder layer, the decoder inserts a third module, which performs Multi-head attention over the output of the encoder stack. We also modify the self-attention sub-layer in the decoder stack to prevent positions from attending subsequent positions. This masking ensures that the predictions for position $i$ can depend only on the known outputs at positions less than $i$
\end{itemize}

In the model, the encoder receives the sequence of the provenance graph as input $(X_1,\ldots, X_n)$, where $X_i {\in} {{\mathbb R}^m}$ denotes the $i$th node representation vector and $n$ represents the length of a provenance graph. Its output matrix Output $z = (Z_1,\ldots, Z_n)$, where $Z_i {\in} {{\mathbb R}^d}$ contains the feature sequence composed of the feature information of each location in different subspaces, and $d$ is the dimension set by the model encoder. In this paper, we select the average vector $Feature {\in} {{\mathbb R}^{d}}$ of the encoder's output matrix $Output$ as a representation of the whole sequence, which can comprehensively take into account the information of the whole sequence to be used for the following classification and detection steps.

There are two inputs in the decoder, one is the output of the Encoder, and another is the features generated with the normalized summation of the output. The Decoder utilizes multi-head attention on the Encoder's output to learn how to reproduce the input sequence based on the feature retrieved by the Encoder. It generates an output sequence $(Y_1,\ldots, Y_n)$ as the reproduced provenance graph feature sequence at a time. At each step the model is auto-regressive\cite{2013Generating}, consuming the previously generated sequences as additional input when generating the next. 

The model is trained by minimizing the error between the input feature sequence and the reconstructed feature sequence. We choose the MSE loss function and apply the Adam optimizer to train the model by minimizing the reconstruction error.
The loss function is defined as shown in the following formula:
\begin{equation}
loss = {\|X-X^{'}\|}_F.
\end{equation}

Due to the difficulties of gathering attack data and modeling attack behavior to cover all attack modes in the security area, it is not suitable to use the system's normal behavior data and attack data to train the model supervised at the same time. This paper decides to train using only normal data. 

We extract the long-term features of the provenance graph sequence using the model's encoder in the long-term feature extraction stage.  The Encoder module receives the feature sequence $X$ of each provenance graph as input, and the output includes the feature sequence $Output$ and the feature vector $Feature$, where $Output {\in} {{\mathbb R}^{n \times d}}$ is a sequence composed of a series of feature vectors, whose dimension is equal to the $d_{model} = 512$ sets by the model, and the length of the sequence is equal to the length of the input sequence. We choose the normalized summation of the $Output$ as the $Feature$, which can comprehensively take into account the information of the whole sequence. As shown in the following formula:
\begin{equation}
    Output, Feature = model.Encoder(X), 
\end{equation}
where $X {\in} {{\mathbb R}^{n \times m}}$ is the input sequence.

Each feature vector in the $Output$ contains the relationship between the information of this position and other positions, thus capturing the long-term dependencies of a provenance graph. To effectively cluster and classify the long-term features extracted from the provenance graph, we normalize and sum the $Output$ to obtain a one-dimensional provenance graph feature vector $Feature$. The $Feature$ vector incorporates all the features of the entire sequence and can effectively reflect the long-term historical information of the provenance graph. Since the system's behaviors are captured in these feature vectors, similar behaviors will cluster together in the embedding space. As a result, we divide all of the feature vectors in the training set into K classes using the clustering algorithm K-means\cite{1979Algorithm}, and the $K$ clustering centers serve for later attack detection.

{\it LTRDetector} uses unsupervised one-class learning, which simply calls for normal system operations, to overcome the limits of binary classification. Each process node's Long-Term implicit relationship is first learned by the encoder, and from the hidden representation, the decoder learns to reconstruct the original node embedding. In the training data set, which only contains normal origin graphs (i.e., unsupervised learning), the objective of the training is to minimize the reconstruction loss.

\section{Experiment}\label{experiment}

\subsection{Experimental Datasets}

To evaluate our method, we employ {\it LTRDetector} on five public datasets, the StreamSpot, two groups of DARPA (Defense Advanced Research Projects Agency, DARPA) TC (Transparent Computing, TC), and two groups of UNICORN SC (Supply Chain).

The StreamSpot\cite{DataStreamSpot} dataset contains 6 scenarios, five of which are normal scenarios (Youtube, Gmail, etc.) and one of which is an attack. In the attack scenario, a program is downloaded from a malicious URL and a flash memory vulnerability is exploited to gain system administrator privileges. This dataset was gathered by running each scenario 100 times using the Linux SystemTap logging system. As a result, it generated 600 execution logs (i.e., 600 provenance graphs). More detailed information about each provenance graph is shown in Table\ref{tab:streamspot}.

The DARPA TC comes from the Defense Advanced Research Projects Agency (DARPA) Transparent Computing (TC) program. In addition to gathering detailed behavioral information from systems, performing attack detection and forensic provenance, the project arranges red and blue teams to carry out offensive and defensive exercises. The CADETS dataset was gathered on FreeBSD (Causal, Adaptive, Distributed, and Efficient Tracing System). The ClearScope data was gathered on Android. More detailed information about each provenance graph is shown in Table\ref{tab:TC}.
\begin{table}[!t]
    \caption{Characteristics of the StreamSpot dataset\label{tab:streamspot}}
    \centering
    \renewcommand{\arraystretch}{1.2}
        \begin{tabular}{p{1.3cm}<{\centering}|c|c|c|c|p{1.3cm}<{\centering}}
        \hline
        \multirow{2}{*}{Dataset}     & \multirow{2}{*}{Label}     & \multirow{2}{*}{Graphs}     & \multirow{2}{*}{Avg.$|$V$|$}  & \multirow{2}{*}{Avg.$|$E$|$}     & Data Size (GiB)    \\
        \hline
        \multirow{6}*{StreamSpot} & YouTube   & 100        & 8292         & 113229           & 0.3               \\
        \cline{2-6}
        ~        & YouTube   & 100        & 8292         & 113229           & 0.3               \\
        \cline{2-6}
        ~        & Gmail     & 100             & 6827   & 37382      & 0.1                     \\
        \cline{2-6}
        ~        & Download  & 100             & 8831   & 310814     & 1                       \\
        \cline{2-6}
        ~        & VGame     & 100             & 8637   & 112958     & 0.4                     \\
        \cline{2-6}
        ~        & CNN       & 100             & 8990   & 294903     & 0.9                     \\
        \cline{2-6}
        ~        & Attack    & 100             & 8891   & 28412      & 0.1                     \\
        \hline
        \end{tabular}
\end{table}

\begin{table}[!t]
    \caption{Characteristics of the DARPA TC dataset\label{tab:TC}}
    \centering
    \renewcommand{\arraystretch}{1.2}
        \begin{tabular}{c|c|c|c|c|p{1.3cm}<{\centering}}
        \hline
        \multirow{2}{*}{Dataset}     & \multirow{2}{*}{Label}     & \multirow{2}{*}{Graphs}     & \multirow{2}{*}{Avg.$|$V$|$}  & \multirow{2}{*}{Avg.$|$E$|$}     & Data Size (GiB)   \\
        \hline
        \multirow{2}*{ClearScope}  & Benign   & 43             & 2309    & 4199309     & 441     \\
        \cline{2-6}
        ~                          & Attack   & 51             & 11769   & 4273003     & 432      \\
        \hline
        \multirow{2}*{CADETS}      & Benign   & 110            & 59983   & 4811836     & 271     \\
        \cline{2-6}
        ~                          & Attack   & 3              & 386548  & 5160963     & 38      \\
        \hline
        \end{tabular}
\end{table}

\begin{table}[!t]
    \caption{Characteristics of the DARPA TC dataset\label{tab:SC}}
    \centering
    \renewcommand{\arraystretch}{1.2}
        \begin{tabular}{c|c|c|c|c|p{1.3cm}<{\centering}}
        \hline
        \multirow{2}{*}{Dataset}     & \multirow{2}{*}{Label}     & \multirow{2}{*}{Graphs}     & \multirow{2}{*}{Avg.$|$V$|$}  & \multirow{2}{*}{Avg.$|$E$|$}     & Data Size (GiB)   \\
        \hline
        \multirow{2}*{wget}         & Benign   & 125             & 265424    & 975226     & 64     \\
        \cline{2-6}
        ~                           & Attack   & 25              & 257156    & 957968     & 12      \\
        \hline
        \multirow{2}*{shellshock}   & Benign   & 125             & 238338    & 911153     & 59     \\
        \cline{2-6}
        ~                           & Attack   & 25              & 243658    & 949887     & 12      \\
        \hline
        \end{tabular}
\end{table}

The SC dataset was collected by UNICORN\cite{han2020unicorn}. The SC dataset simulated two APT attack scenarios, the wget and shellshock. The experiment was conducted for three days for each scenario, and CamFlow\cite{pasquier2017practical} was employed to gather provenance graphs. Table\ref{tab:SC} displays the statistics for the datasets wget and shellshock.

\subsection{Experimental Setup}

In the experimental stage, the normal data was split into three sections, the training set, the verification set, and the test set which includes all attack data. The division of normal data in this experiment follows the same percentage as the comparison method UNICORN\cite{han2020unicorn}. 25\% of normal data from the StreamSpot\cite{DataStreamSpot} dataset is divided into test sets, 20\% of the wget and shellshock datasets are used for testing, and 10\% of the ClearScope and CADETS datasets. The experiment uses a five-fold cross-validation strategy to reduce the impact of the dataset splitting on the experimental results.

\begin{figure*}[t!]
    \begin{center}
        \begin{subfigure}[t]{0.3\textwidth}
            \centering
            \includegraphics[width=\textwidth]{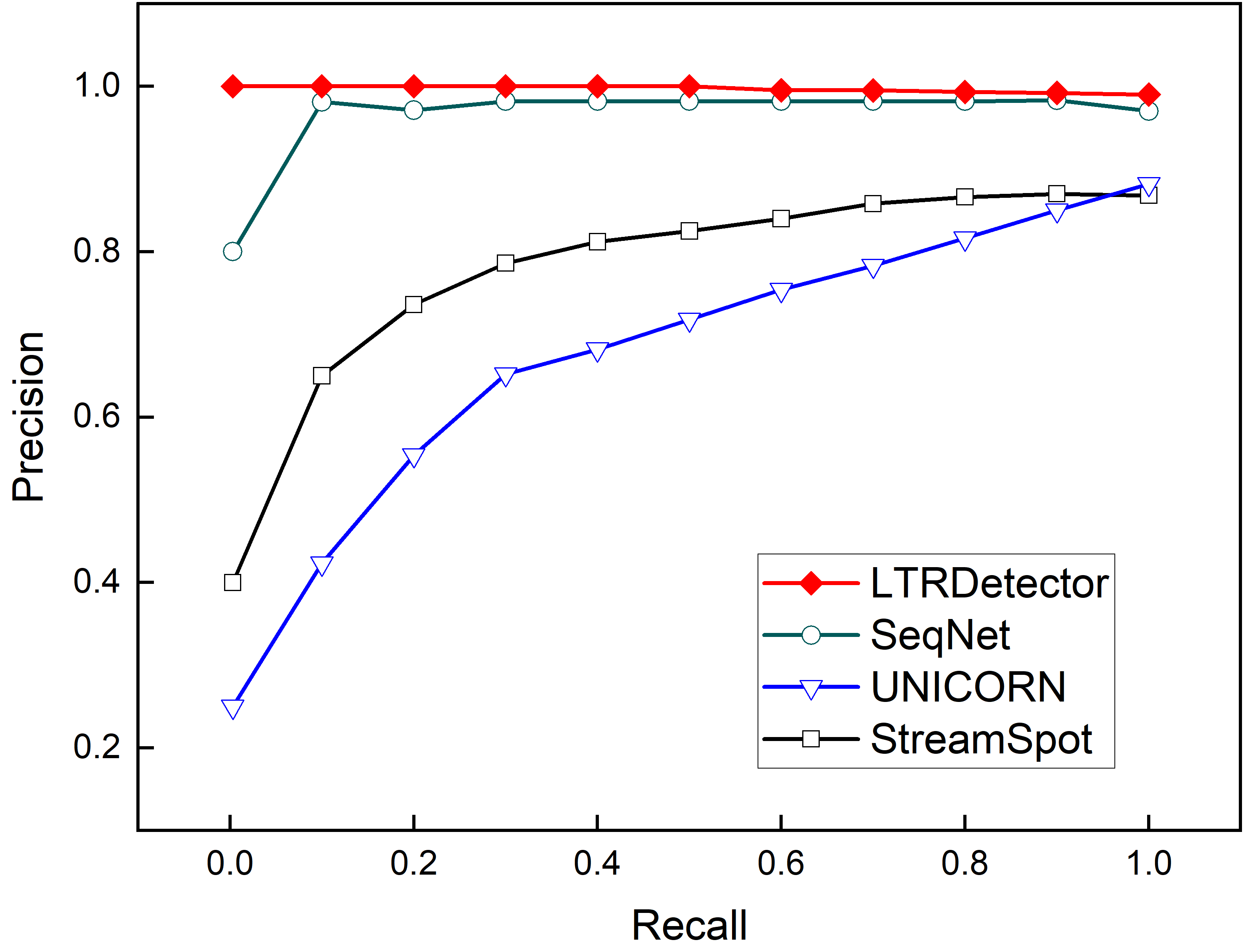}
             \caption{StreamSpot}
     \end{subfigure}
     \begin{subfigure}[t]{0.3\linewidth}
             \centering
             \includegraphics[width=\textwidth]{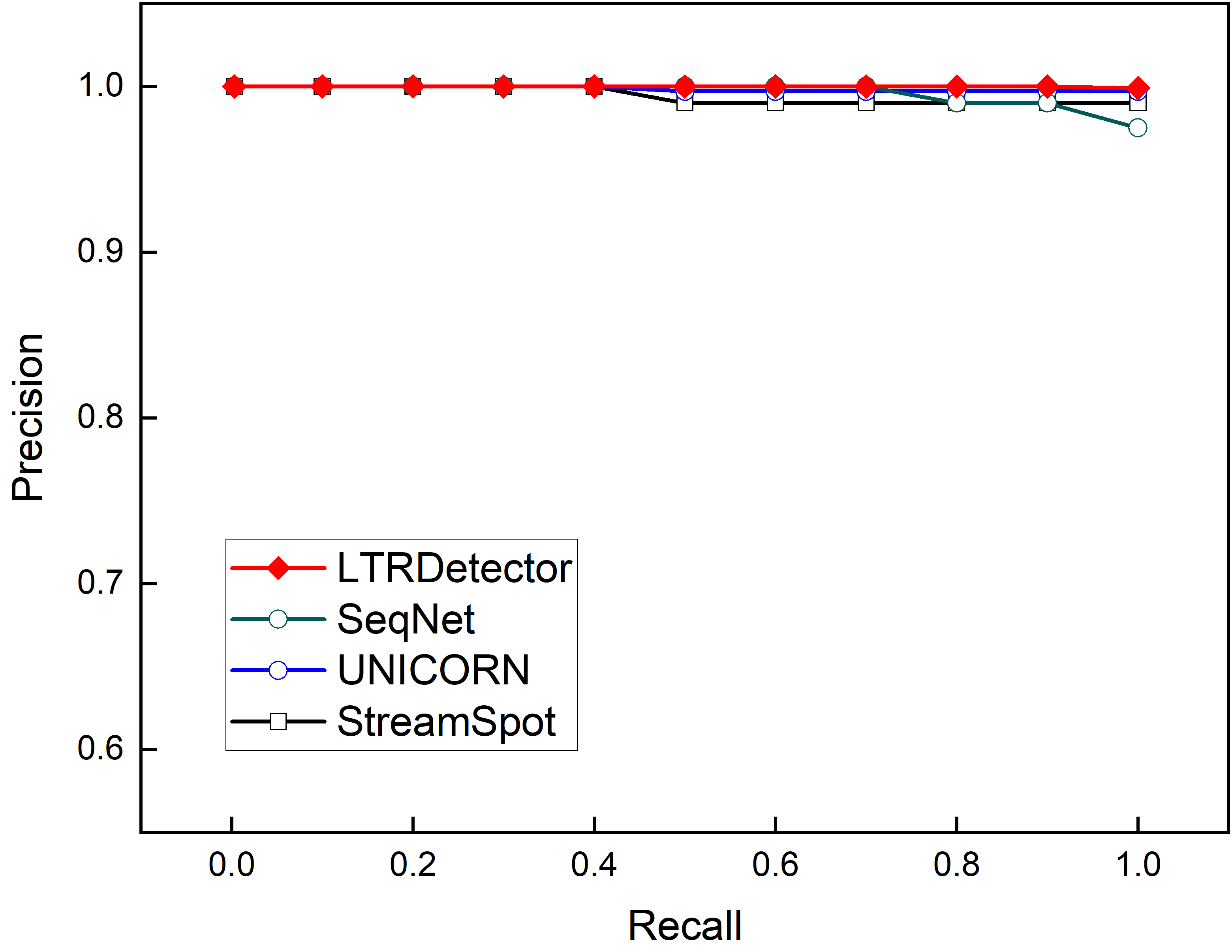}
             \caption{ClearScope}
     \end{subfigure}
     \begin{subfigure}[t]{0.3\textwidth}
             \centering
             \includegraphics[width=\textwidth]{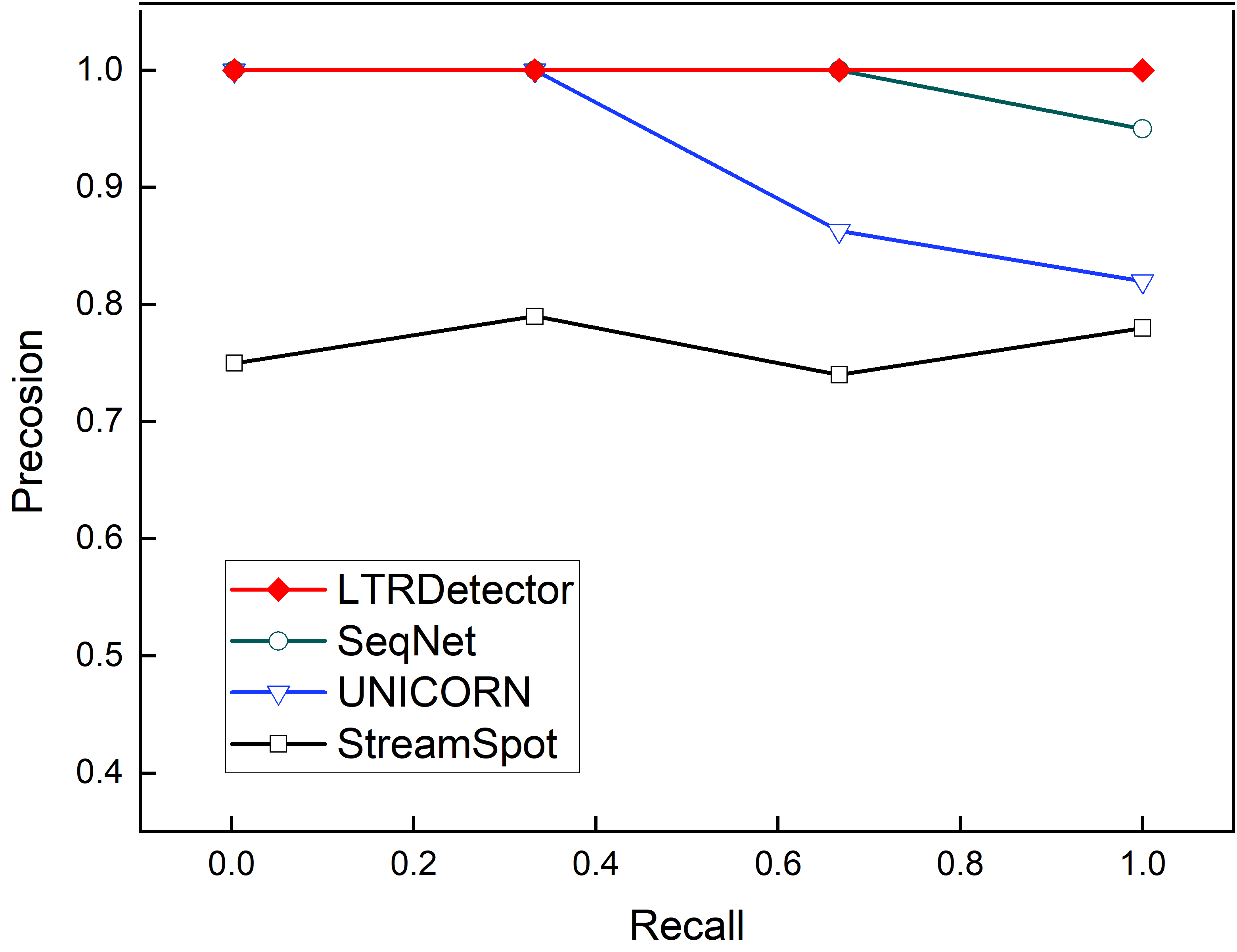}
             \caption{CADETS}
     \end{subfigure}\vspace{0.5cm}
        \begin{subfigure}[t]{0.3\textwidth}
            \centering
            \includegraphics[width=\textwidth]{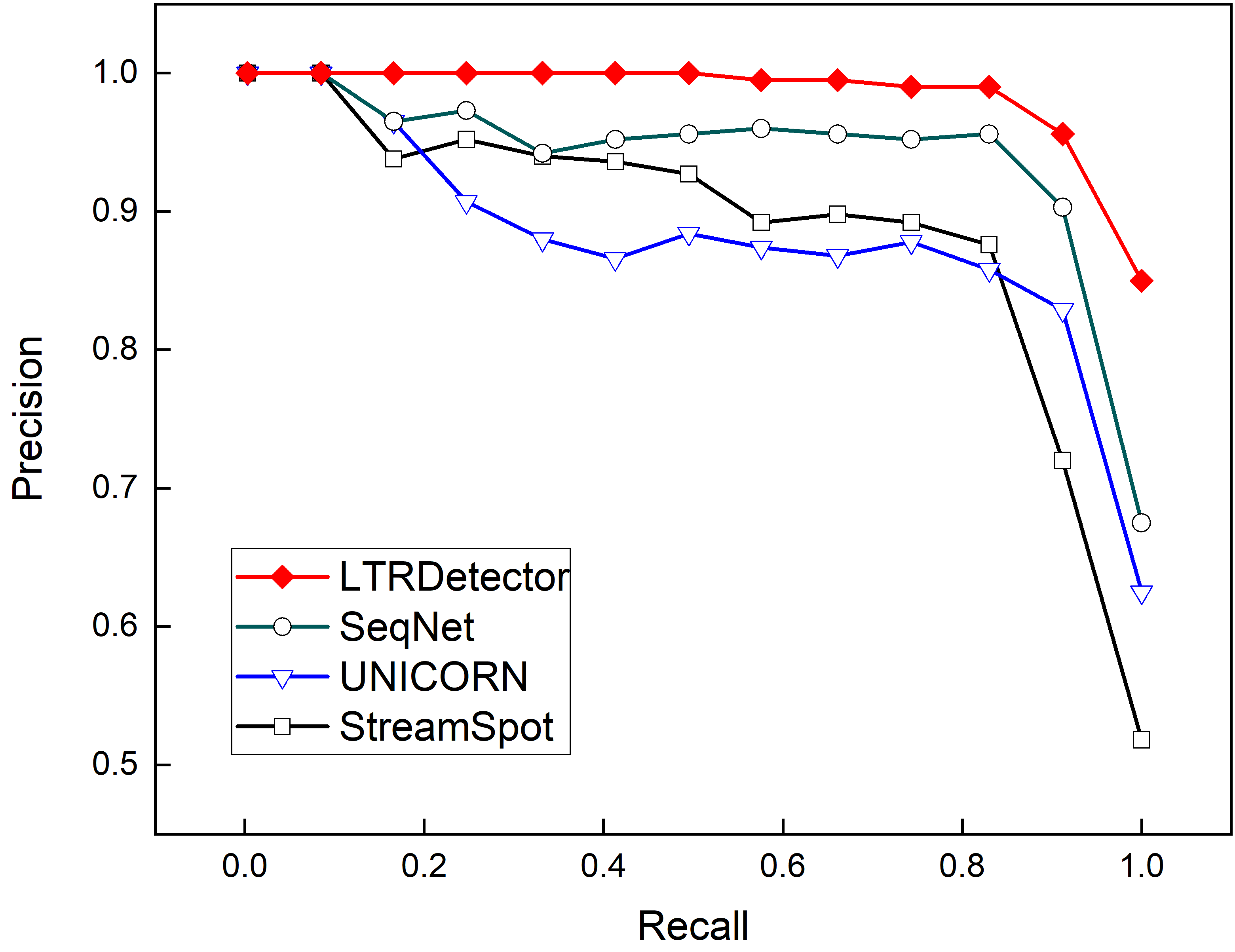}
             \caption{wget}
     \end{subfigure}
     \begin{subfigure}[t]{0.3\textwidth}
        \centering
        \includegraphics[width=\textwidth]{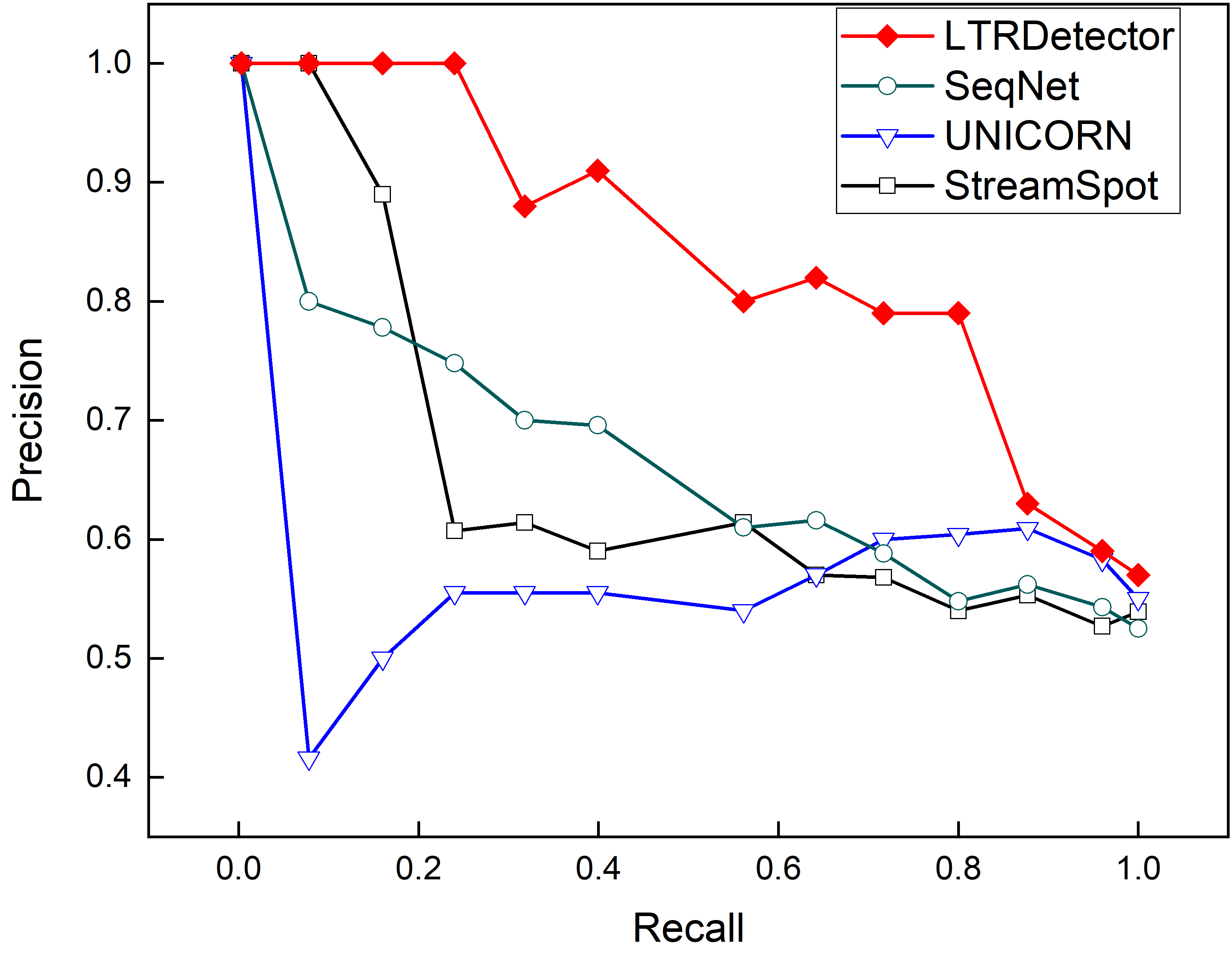}
        \caption{shellshock}
\end{subfigure}
    \end{center}
    \caption{Precision-Recall of comparison experiments on five datasets.}\label{compare}
\end{figure*}


\begin{table*}[!t]
    \caption{Average AUC-PR of comparison experiments\label{tab:compare}}
    \centering
    \begin{tabular}{p{1.5cm}<{\centering} | p{1.3cm}<{\centering} p{1.3cm}<{\centering} p{1.1cm}<{\centering} p{1.3cm}<{\centering} p{1.3cm}<{\centering}}
    \toprule
    Methods  & StreamSpot & ClearScope & CADETS & wget & shellshock \\
    \midrule
    streamspot  & 0.786     & 0.991    & 0.763    & 0.890     & 0.651 \\
    UNICORN     & 0.679     & 0.995    & 0.921    & 0.881     & 0.567 \\
    SeqNet      & 0.968     & 0.994    & 0.988    & 0.942     & 0.664  \\
    LTRDetector & \textbf{0.994}     & \textbf{0.997}    & \textbf{0.997}    & \textbf{0.984}     & \textbf{0.841} \\
    
  \bottomrule
    \end{tabular}
\end{table*}

\begin{figure*}[t!]
    \begin{center}
        \begin{subfigure}[t]{0.3\textwidth}
            \centering
            \includegraphics[width=\textwidth]{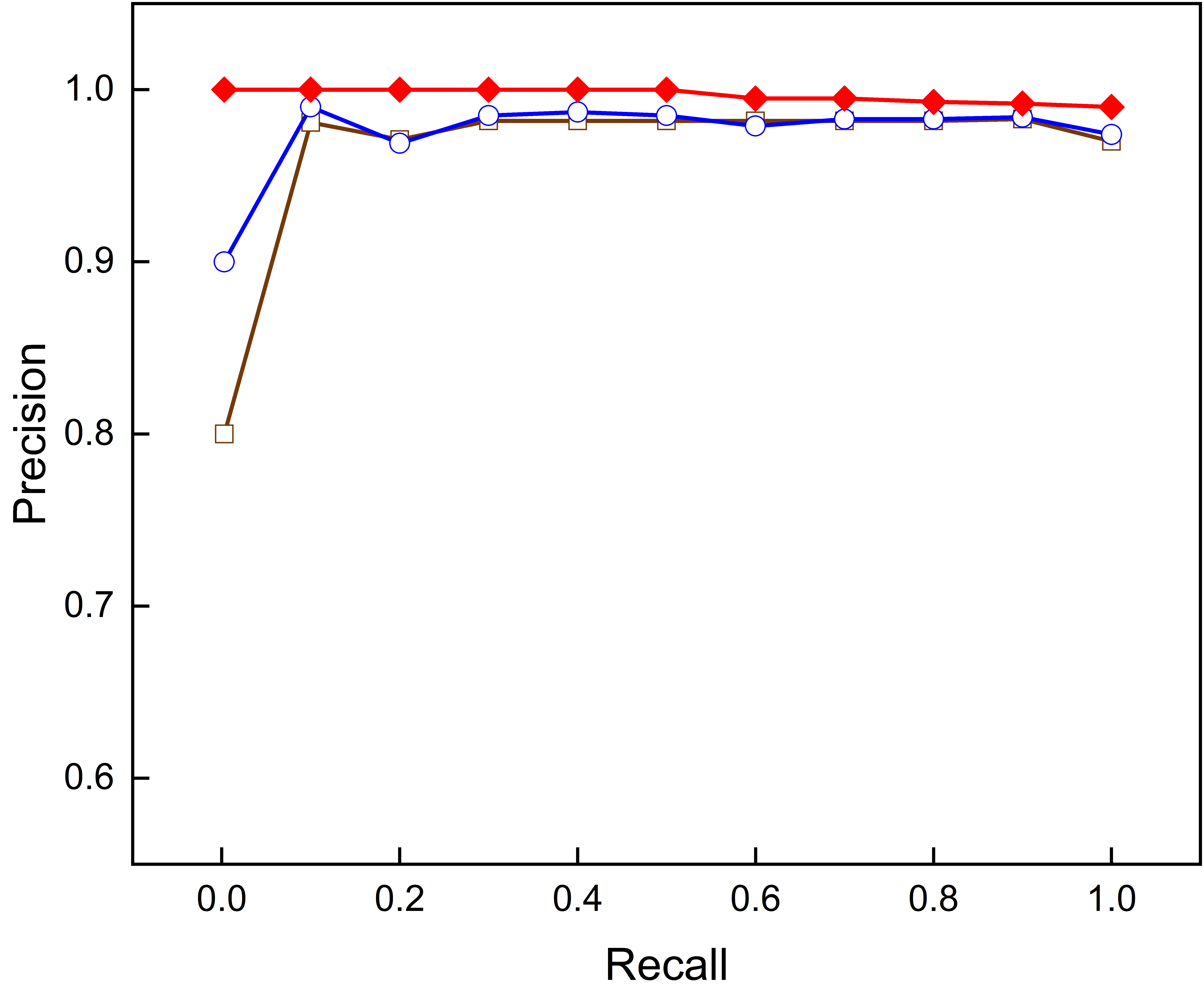}
             \caption{StreamSpot}
    \end{subfigure}
    \begin{subfigure}[t]{0.3\linewidth}
             \centering
             \includegraphics[width=\textwidth]{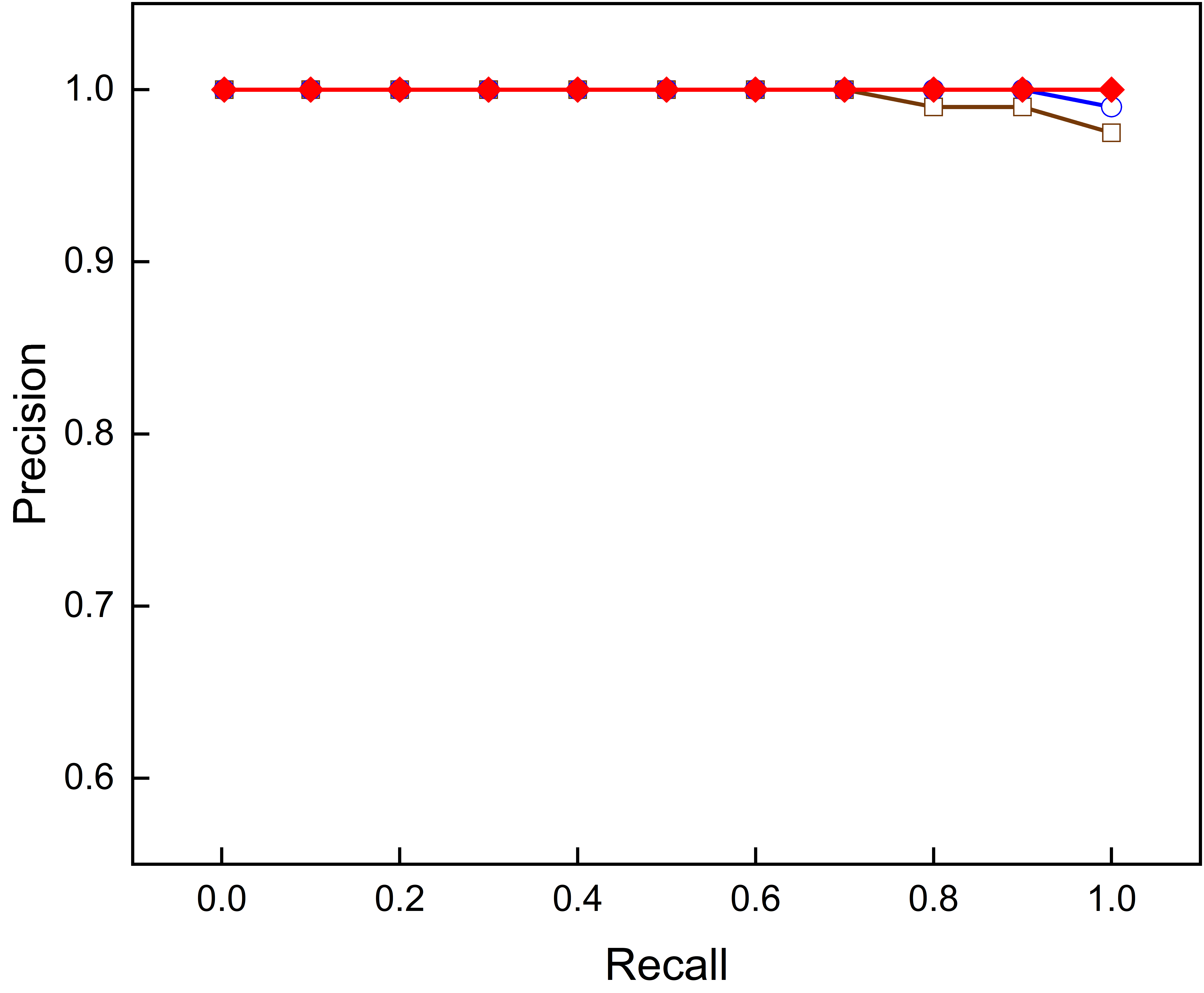}
             \caption{ClearScope}
    \end{subfigure}
    \begin{subfigure}[t]{0.3\textwidth}
             \centering
             \includegraphics[width=\textwidth]{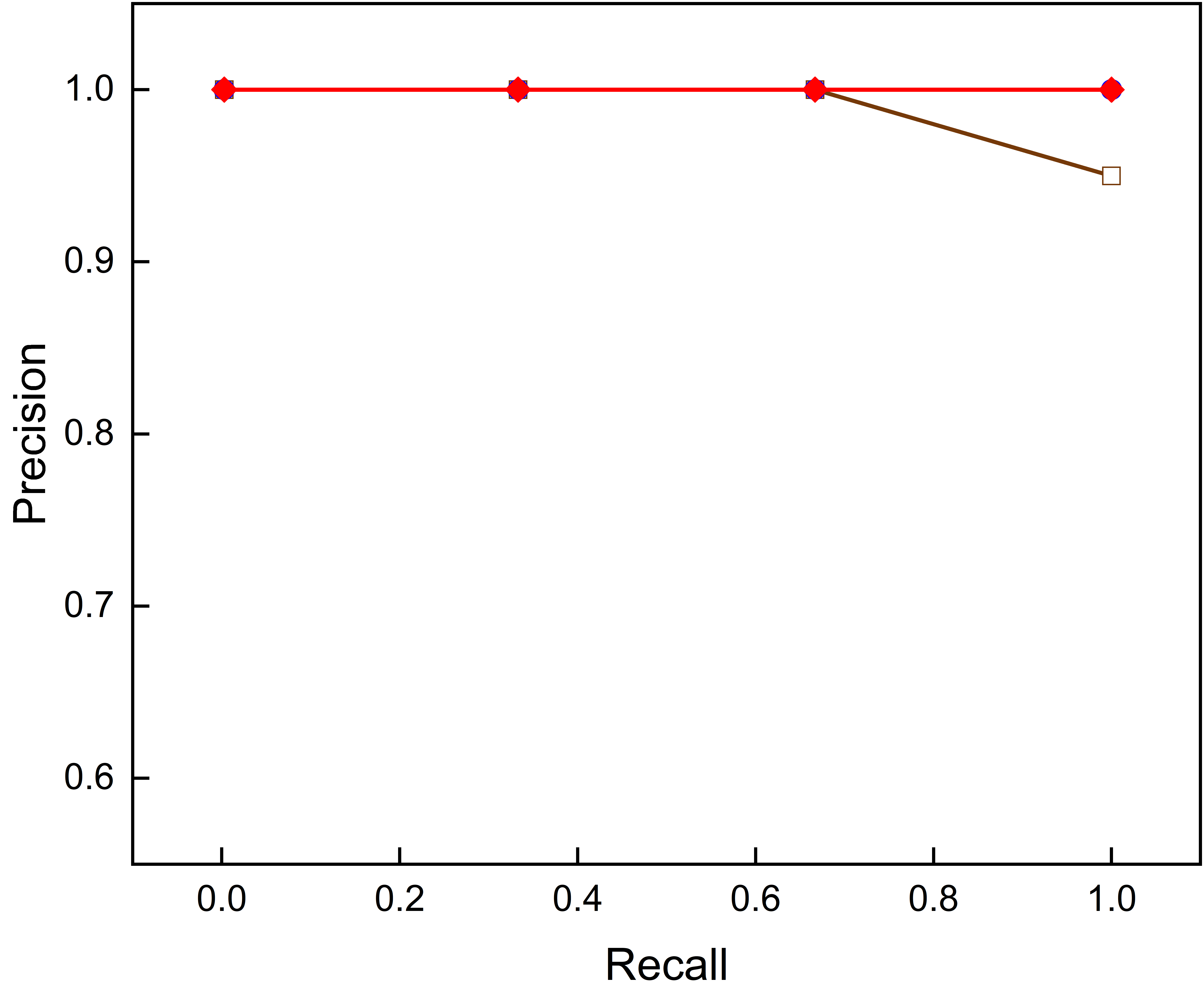}
             \caption{CADETS}
    \end{subfigure}\vspace{0.5cm}
    \begin{subfigure}[t]{0.3\textwidth}
            \centering
            \includegraphics[width=\textwidth]{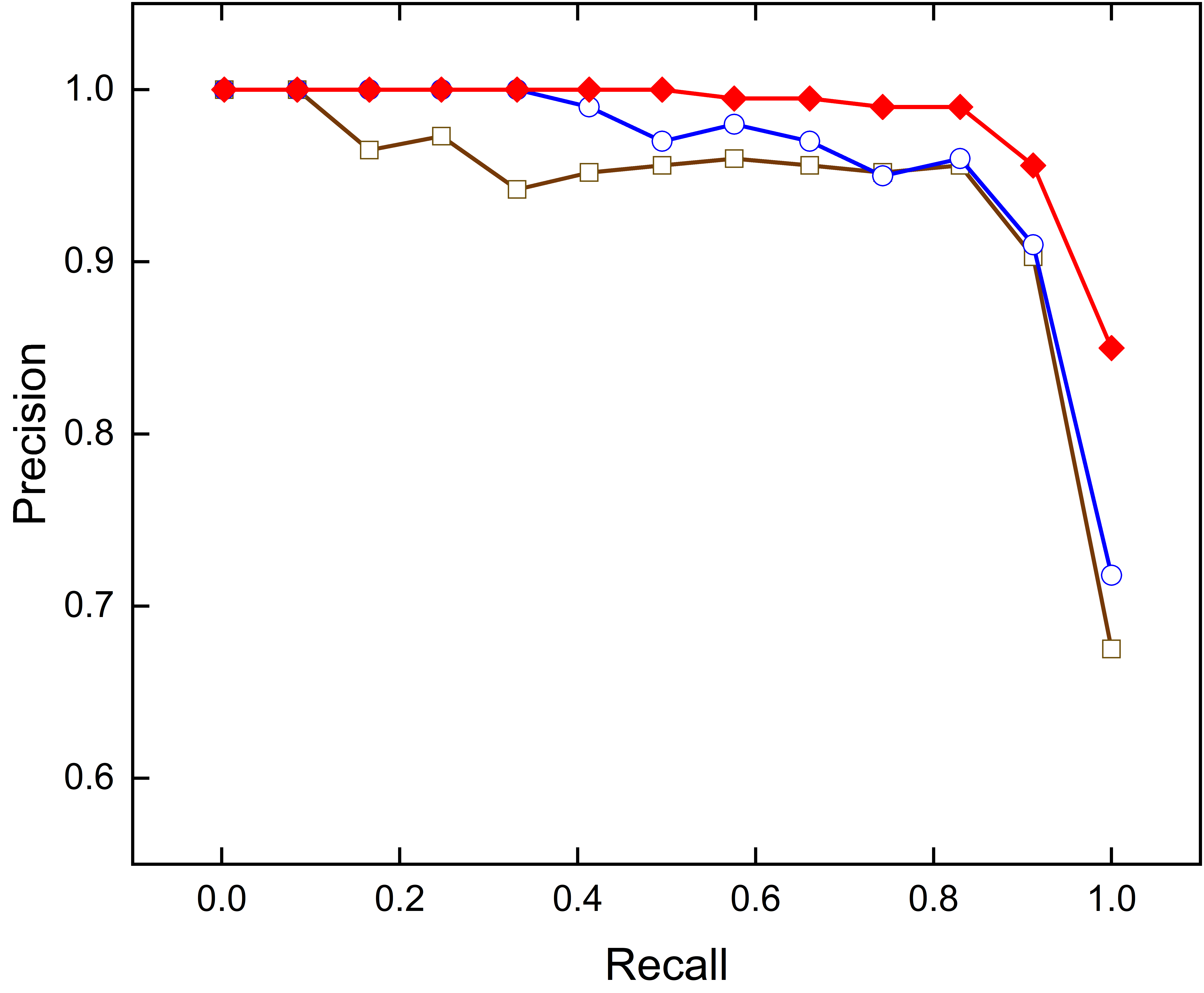}
             \caption{wget}
    \end{subfigure}
    \begin{subfigure}[t]{0.3\textwidth}
        \centering
        \includegraphics[width=\textwidth]{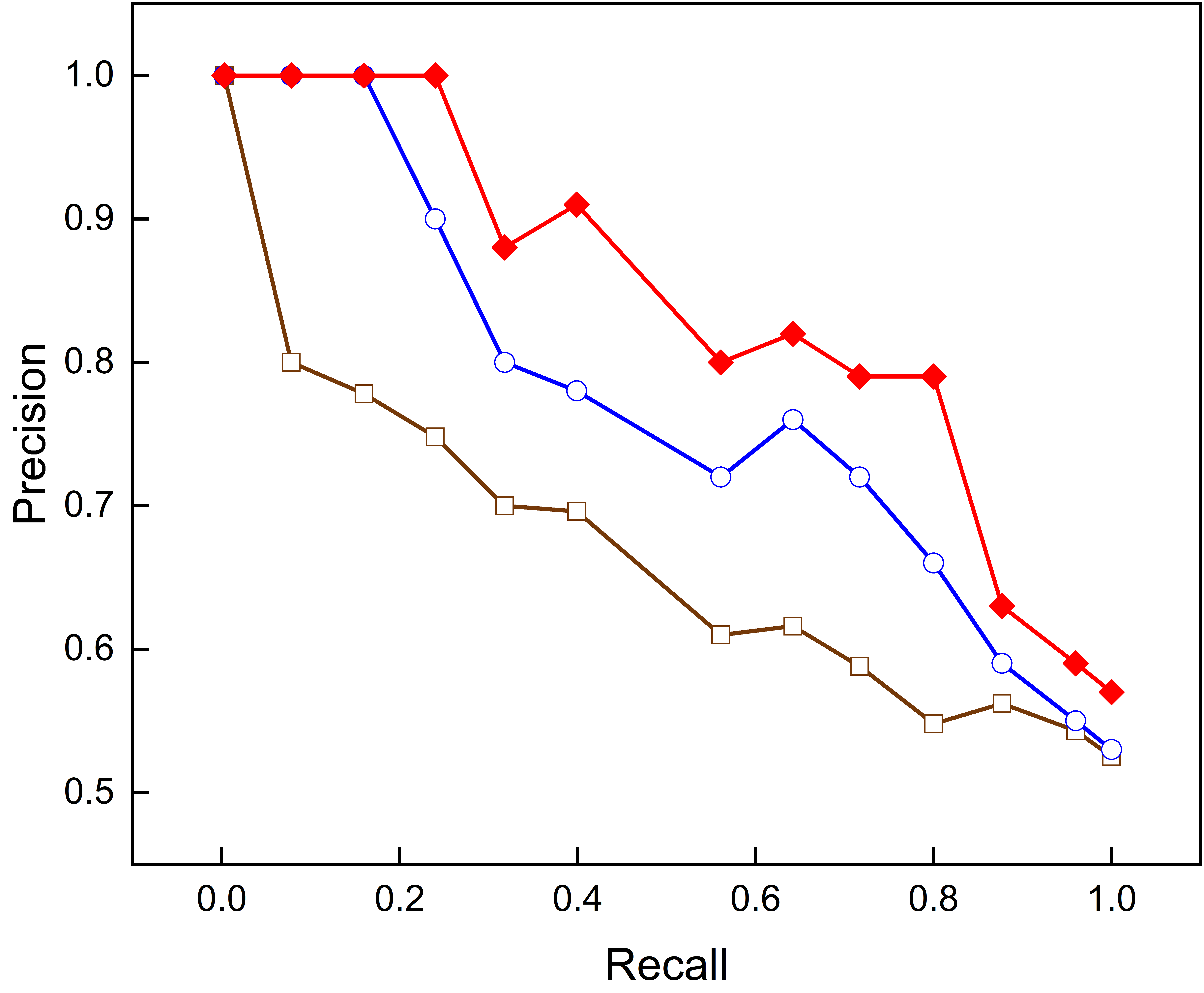}
        \caption{shellshock}
    \end{subfigure}
    \begin{subfigure}[t]{0.80\textwidth}
        \centering
        \includegraphics[width=\textwidth]{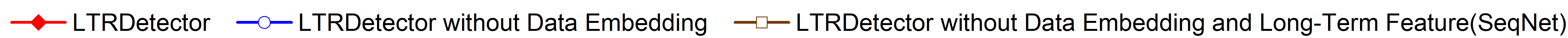}
    \end{subfigure}
    \end{center}
    \caption{Precision-Recall of ablation experiments on five datasets.}\label{ablation}
\end{figure*}

\subsection{Evaluation Criteria}

Since this paper models the normal behavior of the system, it is necessary to specify the threshold value during the testing process. As it is impossible to determine the ideal threshold in a real environment, in this paper, we draw the PR (Precision-Recall) curve based on the estimation of precision and recall at different thresholds. Then we use the area between the PR curve and the coordinate axis as the evaluation criteria (AUC-PR), and a larger area indicates a better overall performance of classification.

Precision and Recall are two major evaluation indexes for classification problems. These two indicators are derived from four sets of information: true positives (TP), false positives (FP), true negatives (TN), and false negatives (FN).

In this study, the Precision is determined using the formula below:
\begin{equation}
Precision = \frac{TP}{TP + FP}.
\end{equation}
Precision indicates the proportion of real attacks among all data that have been classified as attacks.

As follows is the calculation for the Recall:
\begin{equation}
Recall = \frac{TP}{TP + FN}.
\end{equation}
Recall indicates the proportion of detected attacks to all real attacks.

\subsection{Comparison Method}

We compared {\it LTRDetector} with the following three mainstream APT attack detection methods.

StreamSpot\cite{manzoor2016fast}: The StreamSpot approach uses the similarity of local subgraphs to define the similarity between provenance graphs, and then uses a clustering algorithm to group similar provenance graphs together.

UNICORN\cite{han2020unicorn}: UNICORN turns the provenance graph into a histogram, and employs the HistoSketch algorithms\cite{yang2017histosketch} to convert the histogram into feature vectors. Then it creates state nodes by clustering the feature vectors and describes the behavior of the system by analyzing the transfer between state nodes.

SeqNet\cite{2022SeqNet}: SeqNet uses the same provenance graph embedding algorithm as UNICORN\cite{han2020unicorn}, which converts the provenance graph sequences into feature vectors. And then use the GRU model to extract Long-Term features of the provenance graphs.

\subsection{Comparative Experimental Results}

To evaluate our method, we compare {\it LTRDetector} with the current representative approaches SeqNet\cite{2022SeqNet}, UNICORN\cite{han2020unicorn}, and StreamSpot\cite{manzoor2016fast} on five public datasets. The PR curves are shown in Figure\ref{compare}. The recall rate and matching precision are the horizontal and vertical coordinates of the PR curve. We use the area under the PR curve as the average AUC-PR of each approaches on five datasets as shown in Table\ref{tab:compare}.
\begin{enumerate}
    \item Compared with the APT detection method of StreamSpot, {\it LTRDetector} outperforms the compared methods on five datasets consistently. The improvements are 20.92\% on the StreamSpot dataset, 0.6\% on the ClearScope dataset, 23.47\% on the CADETS dataset, 9.55\% on the wget dataset, and 22.59\% on the shellshock dataset. 
    \item Compared with the UNICORN APT detection method, the improvements are 31.69\% on the StreamSpot dataset, 0.2\% on the ClearScope dataset, 7.62\% on the CADETS dataset, 10.47\% on the wget dataset, and 32.58\% on the shellshock dataset.
    \item Compared with the SeqNet APT detection method, the improvements are 2.62\% on the StreamSpot dataset, 0.3\% on the ClearScope dataset, 0.9\% on the CADETS dataset, 4.27\% on the wget dataset, and 21.05\% on the shellshock dataset.
\end{enumerate}


\begin{table*}[!t]
    \caption{Average AUC-PR of ablation experiments\label{tab:ablation}}
    \centering
    \begin{tabular}{p{5.5cm}<{\centering} | p{1.3cm}<{\centering} p{1.3cm}<{\centering} p{1.1cm}<{\centering} p{1.3cm}<{\centering} p{1.3cm}<{\centering}}
    \toprule
    Methods  & StreamSpot & ClearScope & CADETS & wget & shellshock \\
    \midrule
    w/o Data Embedding and Long-Term Feature & 0.968 & 0.994 & 0.988 & 0.942 & 0.664 \\
    w/o Data Embedding  & 0.975 & 0.996 & \textbf{0.997} & 0.962 & 0.774  \\
    LTRDetector  & \textbf{0.994} & \textbf{0.997} & \textbf{0.997} & \textbf{0.984} & \textbf{0.841} \\
    
  \bottomrule
    \end{tabular}
\end{table*}

From Figure\ref{compare}, we can see that our approach outperforms SeqNet\cite{2022SeqNet}, UNICORN\cite{han2020unicorn}, and StreamSpot\cite{manzoor2016fast} in this investigation. There are primarily two factors for this. First, we employ the Word2Vec\cite{mikolov2013distributed} as the embedding technique to successfully retain network topology and contextual information. Furthermore, we apply the Multi-head attention\cite{vaswani2017attention} algorithm, which is capable of efficiently extracting the Long-Term features, allowing it to better adapt to APT attacks in real-life scenarios. 

The advantage of {\it LTRDetector} over the StreamSpot method is that the sequence of provenance graphs is extracted into a sequence of feature vectors, which contains the characteristics of system state changes. The UNICORN method uses feature sequences as well, but it ignores small differences between anomalous and normal behavior, which makes it ineffective in detecting long-period attack activities. Though SeqNet uses the GRU model to extract Long-Term features, it is an RNN model with typical recurrent neural network features, which cannot solve the gradient disappearance and Long-Term dependency problem\cite{bengio1994learning}. SeqNet is therefore unable to successfully extract Long-Term features of long-period APT attacks. In this paper, Node2Vec used as the embedding method incorporates data reduction algorithms CPR\cite{2016High} to retain as much semantic information as possible while reducing the data size. A set of distinctive, independent, and information-rich traits are extracted. In the process of Long-Term features extraction, historical attributes, and current attributes are combined to accumulate small changes in the sequence and gradually widen the gap between normal and abnormal behavior in the feature space. Therefore, our method can detect long-period APT attacks effectively.

\subsection{Ablation experiments}

In this paper, we introduced two steps in our method, one is data processing and the other is Long-Term feature extraction. The efficacy of each step is shown in Figure\ref{ablation}. We verify the effectiveness of each step by the control variable method. The PR curve of each model is obtained by adding each step to the model, and the area under the PR curve is used as the average AUC-PR of each method on five datasets, as shown in Table\ref{tab:ablation}.
\begin{enumerate}
    \item The first two groups prove the effectiveness of Long-Term features. The model with Long-Term features outperforms the model without Long-Term features on five datasets consistently. The improvements are 0.71\% on the StreamSpot dataset, 0.16\% on the ClearScope dataset, 0.84\% on the CADETS dataset, 2.08\% on the wget dataset, and 16.66\% on the shellshock dataset. 
    \item The second and third groups prove the effectiveness of data embedding. The model with effective data embedding shown in group three outperforms the model without data embedding on five datasets consistently, the improvements are 2.63\% on the StreamSpot dataset, 2.09\% on the ClearScope dataset, 0.84\% on the CADETS dataset, 4.44\% on the wget dataset, and 26.73\% on the shellshock dataset.
\end{enumerate}

From the above experimental results, it can be seen that the newly added steps all have a corresponding positive impact on the precision of APT attack detection.


\begin{figure*}[t!]
    \begin{center}
        \begin{subfigure}[t]{0.19\textwidth}
            \centering
            \includegraphics[width=\textwidth]{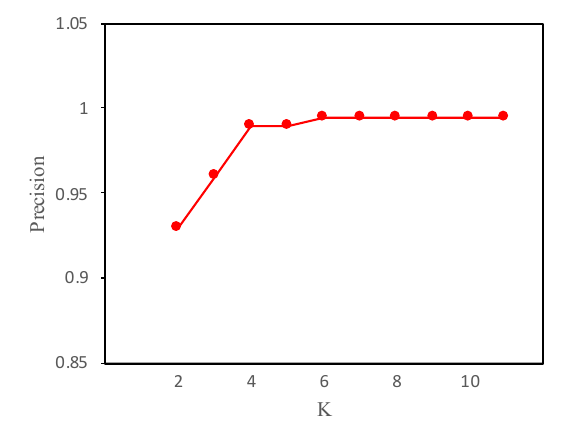}
             \caption{StreamSpot}
    \end{subfigure}
    \begin{subfigure}[t]{0.19\linewidth}
             \centering
             \includegraphics[width=\textwidth]{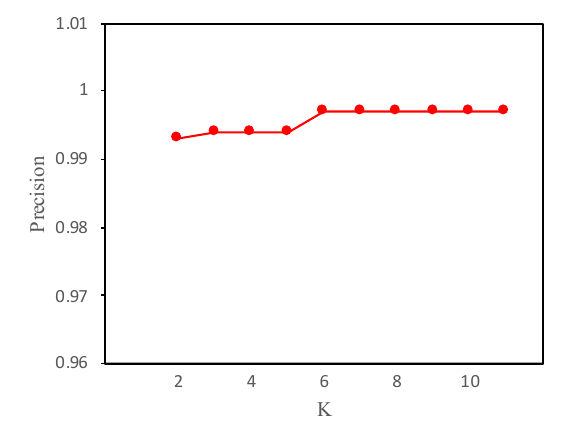}
             \caption{ClearScope}
    \end{subfigure}
    \begin{subfigure}[t]{0.19\textwidth}
             \centering
             \includegraphics[width=\textwidth]{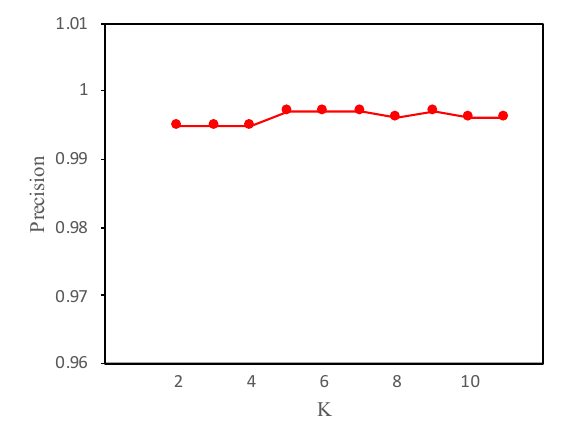}
             \caption{CADETS}
    \end{subfigure}
    \begin{subfigure}[t]{0.19\textwidth}
            \centering
            \includegraphics[width=\textwidth]{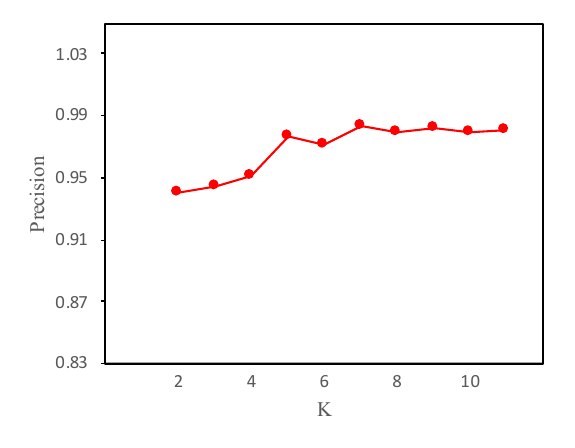}
             \caption{wget}
    \end{subfigure}
    \begin{subfigure}[t]{0.19\textwidth}
        \centering
        \includegraphics[width=\textwidth]{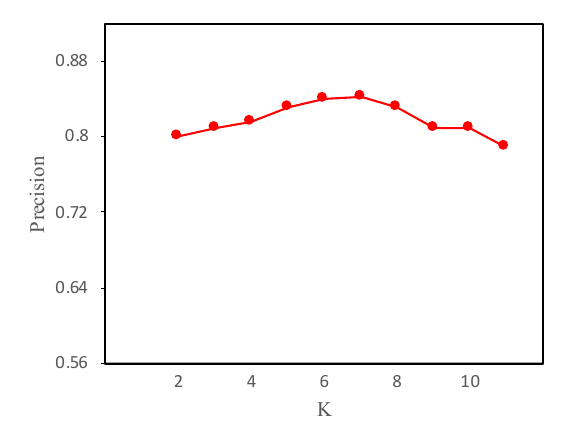}
        \caption{shellshock}
    \end{subfigure}
    \end{center}
    \caption{Average AUC-PR of different cluster center K on five datasets.}\label{param}
\end{figure*}

\begin{table*}[h]
\centering
  \caption{Runtime evaluation}\label{run}
  \begin{tabular}{p{1.5cm}<{\centering} | p{1.3cm}<{\centering} p{1.3cm}<{\centering} p{1.1cm}<{\centering} p{1.3cm}<{\centering} p{1.3cm}<{\centering}}
    \toprule
           & StreamSpot & ClearScope & CADETS & wget & shellshock \\
    \midrule
    streamspot & 4.46s & 42.75s & 5.94s & 16.79s & 32.38s \\
    UNICORN & 0.78s & 0.89s & 0.51s & 2.16s & 1.74s \\
    SeqNet & \textbf{0.23s} & 0.51s & 0.29s & 1.66s & 1.29s \\
    LTRDetector & 0.29s & \textbf{0.49s} & \textbf{0.27s} & \textbf{1.57s} & \textbf{1.22s} \\
  \bottomrule
\end{tabular}
\end{table*}

\subsection{Parameter experiments}

In this paper, we investigate the influence of cluster center number K on detection outcomes by setting different numbers of cluster centers. The P-R curves of the five data sets at different values of K are obtained by setting the number of clustering centers K from 2 to 11 for each data set, and the area under the P-R curve corresponding to the K value is calculated as the average detection accuracy. The experimental results are shown in Figure\ref{param}. 

The overall results of the five data sets show that when the number of clustering centers K is small, the granularity of clustering is coarser, so the clustering algorithm will gather the feature vectors that are farther away together, potentially resulting in features that belong to different categories into the same category. At the same time, because there are fewer clustering centers, each category will contain more features, resulting in more discrepancies between the data within the class, causing the clustering center to move. Since the clustering centers can no longer properly reflect the common characteristics, the gap between the normal behavioral feature vectors and the clustering centers grows, lowering the accuracy and stability of the clustering and making the model's classification worse. For example, because there are five classes of normal behaviors in the StreamSpot dataset when the number of clustering centers is initially set to K=2, the detection accuracy is only 0.93. However, the overall detection accuracy increases to 0.99 when the value of K reaches 5, and the detection accuracy stabilizes above 0.995 when K$>$5. Only the shellshock dataset shows a slightly reducing trend as the value of K grows in the five experimental datasets, whereas the remaining four datasets exhibit a smooth trend. 

As a result, there is a better classification performance when the value of the clustering center K is taken in the range of 5 to 12.

\subsection{Test run time}
All model training and testing were conducted on an NVIDIA GeForce RTX 3060 GPU. This experiment compares the detection runtime on the testing sets of four models across five datasets. {\it LTRDetector} and SeqNet require loading pre-trained deep learning models. All experimental results are based on individual events, and the specific results are shown in Table\ref{run}.

Observing the experimental results, for all datasets, {\it LTRDetector} has the shortest testing time, which is negligible compared to the attack duration, meeting the real-time requirement. Additionally, it was found that the two methods utilizing deep learning models were more efficient during the testing phase compared to the other two models. This is because the detection cost is primarily concentrated on training the deep learning models during the training phase.

\section{Conclusion and Future Works}\label{conclusion}

In this paper, we propose an APT attack detection method based on the provenance graph. The method first converts provenance graph sequences into feature sequences, then uses the Mulit-Head attention algorithm to capture the Long-Term change characteristics of the sequences. In the model training phase, the Adam optimizer is used to train the model by minimizing the reconstruction error. In the attack detection stage, the trained model is used directly to extract the sequence features. Finally, the clustering method is applied to the training data to represent the typical behavior of the system, and sequences that are far from the cluster centers are recognized as attacks in the attack detection stage. To evaluate the effectiveness of the method, {\it LTRDetector} is compared with three state-of-the-art approaches on five widely used datasets and gets improved detection outcomes.

Currently, {\it LTRDetector} models the normal behavior of the system during training, and we do not change the model thereafter to avoid attacks from poisoning the model. In the following work, we will conduct further research on adaptive learning of model updating so that the model can be updated without being poisoned. Meanwhile, we utilize the K-means clustering technique for attack detection. However, K-means may be unable to handle some complicated data distributions for its simplistic assumption of data distribution, leading to subpar classification. To further increase the model's detection accuracy, it is necessary to examine more appropriately tailored detection approaches in the next work. 


\bibliographystyle{IEEEtran}
\bibliography{LTRDetector}

\vfill

\end{document}